\documentstyle[aps,prl,epsbox]{revtex}

\draft

\newcommand{\bra}{\left\langle}
\newcommand{\ket}{\right\rangle}

\newcommand{\fe}{e^{({\rm F})}}
\newcommand{\be}{e^{({\rm B})}}
\newcommand{\se}{e^{({\rm S})}}

\newcommand{\fL}{L^{({\rm F})}}
\newcommand{\bL}{L^{({\rm B})}}
\newcommand{\sL}{L^{({\rm S})}}

\newcommand{\flambda}{\lambda^{({\rm F})}}
\newcommand{\blambda}{\lambda^{({\rm B})}}
\newcommand{\slambda}{\lambda^{({\rm S})}}

\newcommand{\fLambda}{\Lambda^{({\rm F})}}
\newcommand{\bLambda}{\Lambda^{({\rm B})}}

\newcommand{\ftheta}{\theta^{({\rm F})}}
\newcommand{\btheta}{\theta^{({\rm B})}}

\newcommand{\rev}{{\cal R}}

\begin{document}
\draft

\title{A non-equilibrium equality in Hamiltonian chaos}

\author{Shin-ichi Sasa}

\address{
Department of Pure and Applied Sciences, 
          University of Tokyo, \\
         Komaba, Meguro-ku, Tokyo 153, Japan}

\date{June 24, 2000}

\maketitle

\begin{abstract}
We numerically study a billiard system with a time-dependent force,
and our results suggest the existence of a limitation on possible
transitions between steady states in Hamiltonian chaos, in analogy to 
the limitation on transitions between equilibrium states described 
by the second law of thermodynamics. 
This limitation is expressed in terms of irreversible information loss,
which is defined for each trajectory through Lyapunov analysis. 
As a key step in the study, we demonstrate 
a non-equilibrium equality which means that the average of 
the inverse exponential of the irreversible information loss 
is unity, where the  average is taken over initial conditions 
sampled from  the microcanonical ensemble. 
\end{abstract}




The second law of thermodynamics is a formalization of
a fundamental limitation on possible transitions among
equilibrium states \cite{Lieb}.
Although the validity of the second law has been confirmed conclusively 
without exception for more than a century, there yet exists  no  clear 
understanding of how this limitation emerges from purely Hamiltonian systems. 

When the dynamical system in question possesses a mixing property, 
time correlations of  dynamical variables decay. This decay 
characterizes the directional evolution in dynamical systems. 
For this reason, it may be natural to believe 
that the chaotic behavior of a system is related to the 
limitation on possible transitions between its steady states. 
However, as far as we know, 
such a description has never been formulated explicitly.

In this Letter, we study  non-steady behavior of Hamiltonian chaos
resulting from a change in  the value of a  parameter during 
a finite time interval. 
We define a quantity $I$, called  the `irreversible information 
loss',  for each 
trajectory, and demonstrate that it satisfies the non-equilibrium 
equality
\begin{equation}
\bra \exp(-I) \ket_{\rm mc}=1,
\label{main}
\end{equation}
where $\bra \cdot \ket_{\rm mc}$ denotes the average over initial 
conditions sampled from the microcanonical ensemble on an energy 
surface specified initially.  Through the Jensen inequality,
$
\bra \exp(-I) \ket_{\rm mc} \ge  \exp(- \bra I \ket_{\rm mc}),
$
the inequality  
\begin{equation}
\bra I \ket_{\rm mc} \ge 0
\label{ineq}
\end{equation}
is derived. This may be regarded as a representation of
the limitation on possible transitions.


Our search for a non-equilibrium equality of the form (\ref{main}) 
for Hamiltonian chaos was initially motivated by its obvious 
relation to the Jarzynski equality \cite{Jarz,Crooks} 
\begin{equation}
\bra \exp(-\beta (W-\Delta F)) \ket_{\rm c}=1,
\label{Jarz}
\end{equation}
where $W$ is the work done by an external agent, $\Delta F$ is the 
free energy increment for a state transition, and  
$\bra \cdot \ket_{\rm c}$ is the average over all possible 
histories, each of whose weight is determined by the probabilities
of equilibrium fluctuations of the system 
in an isothermal environment with inverse temperature $\beta$. 
We note that the minimum work principle 
$\bra W \ket_{\rm c} \ge \Delta F$
can be derived as the result of (\ref{Jarz}). Considering the 
similarity of (\ref{main}) and (\ref{Jarz}), it is natural to 
call  (\ref{main}) 'a Jarzynski-type equality in Hamiltonian chaos'.  


Although the analysis we give here can be applied to a fairly general
class of Hamiltonian systems,  we focus on  a billiard system with a
time-dependent external  force.  Here, each phase space point 
$\Gamma$ is 
specified by the canonical coordinates $(q_1,q_2,p_1,p_2)$, and the
Hamiltonian we study is given by 
\begin{equation}
H(\{q_i,p_i\};f)={1 \over 2}(p_1^2+p_2^2)+ 
{1 \over d} (r(q_1)^2+q_2^2)^{d/2}+f q_1,
\label{model}
\end{equation}
with 
\begin{equation}
r(q_1)=\theta(q_1){\rm max}(q_1-a,0)+\theta(-q_1){\rm min}(q_1+a,0),
\end{equation}
where $\theta$ is the Heaviside step function. The quantity 
$f$ in (\ref{model}) is an external force and the potential 
represents a soft
interaction with a stadium-shape wall. In this Letter, we set
$(d,a)=(8.0,0.5)$. We numerically solved the evolution equations
\begin{eqnarray}
{d q_i \over dt} &=& p_i, 
\label{qeq}   \\
{d p_i \over dt} &=& -{\partial H \over \partial q_i}
\label{peq}
\end{eqnarray}
by employing the 4-th order symplectic integrator method 
with a time step $\delta t =10^{-3}$.  


We begin our discussion by describing the basic properties of the 
system in the case of a time-independent force, say $f=f_0$. Then, 
the energy is  a constant, say $E_0$. As an example, we consider 
the case in which $E_0=1.0$ and $f_0=0.0$.
Our numerical results for this system lead us to conclude that it
does indeed possesses the mixing property 
with respect to the microcanonical measure. We thus assume that 
a continuous distribution of initial conditions on an energy surface 
evolves into the microcanonical ensemble  in the weak sense when 
the system is left for a sufficiently long time. 


The chaotic nature of the system is explored by applying a 
linear analysis (the 'Lyapunov analysis') to trajectories.
Through the numerical integration of the linearized equations 
of motion corresponding to (\ref{qeq}) and (\ref{peq}),
\begin{eqnarray}
{d \delta q_i \over dt} &=& \delta p_i, 
\label{dqeq}   \\
{d \delta p_i \over dt} &=& 
-\sum_{j} {\partial^2 H \over \partial q_i \partial q_j} \delta q_j,
\label{dpeq}
\end{eqnarray}
we obtained a tangent evolution map from time $s$ to $t$, 
${\cal T}(t,s)$.

To make the discussion more concrete,  we choose a specific 
reference time 
$t_*$ and consider  a set of four orthogonal unit vectors 
$\{ e_i(t_*) \}$ defined in the tangent space at the phase space 
point $\Gamma(t_*)$ at time $t_*$. Then, the vector $e_i(t_*)$
evolves  to  ${\cal T}(t,t_*) e_i(t_*)$  in the tangent 
space  at the phase space point $\Gamma(t)$.
By employing the Gram-Schmidt procedure 
$ {\cal T}(t,t_*) e_i(t_*)$ (where $t \ge t_*$) can be expressed as
\begin{equation}
{\cal T}(t,t_*) {e}_i(t_*) =\sum_{j=1}^{4} 
{e}_j(t)  L_{ji}(t,t_*),
\label{evol:e}
\end{equation}
where $L_{ij}$ is the $(i,j)$ element of an upper triangular matrix
whose diagonal elements are positive, 
and $\{ {e}_i(t) \}$ is the set of four orthogonal unit 
vectors defined in the tangent space at the phase space 
point $\Gamma(t)$.  
The $i$th Lyapunov exponent $\bar \lambda_i$ is calculated as
the long time average of the quantity $\lambda_i(t)$ given by
\begin{equation}
\lambda_i(t)= {d \over dt} \log L_{ii}(t,t_*).
\label{lam}
\end{equation}
Although the vector ${e}_i(t)$ and the quantity $\lambda_i(t)$ 
depend on the choice of the set $\{ e_i(t_*) \}$,  the value of 
$\bar \lambda_i$ is independent of this choice for
the system in question, and we have 
$ \bar \lambda_1=- \bar \lambda_4 > 0 $ and $\bar \lambda_2= 
\bar \lambda_3=0$. 


There is  a special vector $\se_{1}(\Gamma(t);f_0)$ at the point 
$\Gamma(t)$, which is  defined as
\begin{equation}
\se_{1}(\Gamma(t);f_0)= \lim_{t_* \rightarrow -\infty} {e}_1(t).
\label{lv1:def}
\end{equation}
In our numerical experiments, we insure that $\se_{1}(\Gamma(t);f_0)$ 
is represented by $e_1(t)$ by choosing the value of $t_*$ so that
$t-t_*$ is sufficiently large for the condition 
\begin{equation}
|1-({e}_1(t), {e}_1'(t))^2| \le \epsilon
\label{con}
\end{equation}
to be satisfied for two vectors $e_1(t_*)$ and $e_1'(t_*)$ 
chosen randomly, where $\epsilon$ is arbitrarily chosen to be $10^{-6}$. 
The vector $\se_{1}(\Gamma(t);f_0)$ corresponds to the most unstable 
direction and is called  the 'first Lyapunov vector' at the point 
$\Gamma(t)$. Using this vector, we can define the first local 
expansion ratio  as
\begin{equation}
\slambda_{1}(\Gamma(t);f_0)= {d \over dt}  \log \sL_{11}(t,t_*),
\end{equation}
where $\sL_{11}(t,s)$ is the magnitude of the vector
${\cal T}(t,t_*) \se_{1}(\Gamma(t_*);f_0)$. 


One may expect that $\se_{2}(\Gamma(t);f_0)$ and 
$\slambda_{2}(\Gamma(t);f_0)$ can be defined similarly to
$\se_{1}(\Gamma(t);f_0)$ and $\slambda_{1}(\Gamma(t);f_0)$. 
However, numerically, the convergence condition in this case, taking
a form  similar to  (\ref{con}), is  difficult to realize. 
We thus added an artificial, small damping term $-\eta \delta p_i$ to
(\ref{dpeq}) in order to obtain $\se_{2}(\Gamma(t);f_0)$.
We expect the value obtained for  $\se_{2}(\Gamma(t);f_0)$ in this manner
to be well-defined in the limit $\eta \rightarrow 0_+$ after first taking
$t_* \rightarrow -\infty$. We call the subspace spanned by
$\se_{1}(\Gamma(t);f_0)$ and $\se_{2}(\Gamma(t);f_0)$ 
the `semi-unstable space'.


Next, we consider the system with a time-dependent force.
We consider a function  $f(t)$ that changes from $f_0$ to $f_1$ 
only during a finite time interval $[t_0,t_1]$.
The initial conditions at $t=\tau_{-}$ $ (< t_0)$ 
are sampled from the microcanonical ensemble on an energy surface 
with  energy $E_0$.

The Lyapunov analysis for the time-dependent system is developed in the 
following way.  First, using the Gram-Schmidt decomposition of the 
linearized evolution map ${\cal T}(t,s)$, the time evolution of 
a set of orthogonal unit vectors  is determined. However, the Lyapunov vector  
$\se_{i}(\Gamma(t);f_0)$ does not have meaning for $t > t_0$. 
Instead, we define the `forward Lyapunov vector' $\fe_i(t)$ as the 
unit vector at time $t$ obtained from $\se_{i}(\Gamma(\tau_-);f_0)$ 
under the Gram-Schmidt procedure, and the `backward Lyapunov 
vector' $\be_i(t)$ as the unit vector from which 
$\se_{i}(\Gamma(\tau_+);f_1)$ is obtained under the Gram-Schmidt procedure, 
where $t_1 < \tau_+  $. 
Since $\fL_{ii}(t,\tau_-)$ and $\bL_{ii}(\tau_+,t)$ are 
determined simultaneously with $\fe_i(t)$ and $\be_i(t)$
through these Gram-Schmidt procedures, we define the forward  
local expansion ratio $\flambda_i(t)$ as 
\begin{equation}
\flambda_i(t)={d \over dt} \log \fL_{ii}(t,\tau_-),
\end{equation}
and  the backward  local expansion ratio $\blambda_i(t)$ as 
\begin{equation}
\blambda_i(t)={d \over dt} \log \bL_{ii}(\tau_+,t).
\end{equation}
When the force $f$ is time-independent (say $f_0$), three quantities 
$\flambda_i(t)$, $\blambda_i(t)$ and  $\slambda_i(\Gamma(t);f_0)$
are identical. This leads us to conjecture that the difference 
between $\flambda_i(t)$ and $\blambda_i(t)$ provides a useful
characterization of phenomena exhibited uniquely by 
time-dependent systems. Also, the long time average of 
$\slambda_1(\Gamma(t);f_0)$ is known to be equal to 
the information loss rate. With these in mind, 
we define the irreversible information loss $I$ as
\begin{equation}
I=
\lim_{\eta\rightarrow 0_+} 
\lim_{\tau_{\pm} \rightarrow \pm \infty}
\int_{\tau_-}^{\tau_+} dt \sum_{i=1}^2 [\flambda_i(t)-\blambda_i(t)],
\label{Idef}
\end{equation}
where we remark that the sum is taken over indices corresponding to 
the semi-unstable space, not unstable space.


Now, we present  numerical results which indicate that the non-equilibrium 
equality (\ref{main}) holds for the irreversible information loss $I$
defined by (\ref{Idef}). We first note that  $\be_i(t)$ cannot 
be calculated 
directly,  because the backward evolution of the semi-unstable space is 
not numerically stable. Thus, in order to be able to apply the type of
analysis we discuss here  to the general situation, we need to devise
some more sophisticated algorithm to calculate $\be_i(t)$.
In this Letter, however, we focus on the simplest case in which 
$f$ is changed from $f_0$ to $f_1$ instantaneously at $t=t_0$. 
In this case, 
when $t \ge t_0$, $\be_i(t)$ and $\blambda_i(t)$ are equal to 
$\se_i(\Gamma(t);f_1)$ and $\slambda_i(\Gamma(t);f_1)$, 
respectively. That is, the integration over the region 
$t_0 \le  t \le t_+$ in (\ref{Idef}) can be done numerically.  
The other part of integration can be evaluated  with the aid of
the time-reversed trajectory 
$\{\tilde \Gamma(t) \}_{t=-\tau_+}^{-\tau_-}$, which is
the solution to the  Hamiltonian equation with the time-reversed 
external force, $\{\tilde f(t) \}_{t=-\tau_+}^{\tau_-}$,
and with the initial condition $\tilde \Gamma(-\tau_+)
={\cal R} \Gamma(\tau_+)$, where ${\cal R}$ is the operator
that changes the sign of the momentum. 
We can carry out Lyapunov analysis for the time-reversed trajectory,
and we  express  all quantities obtained in this analysis by simply
adding a tilde to the corresponding quantities in the normal Lyapunov 
analysis. 
{}From the reversibility of the Hamiltonian equations, 
the integration over the region $\tau_-\le t \le t_0$ in (\ref{Idef}) 
turns out to be equal to 
\begin{equation}
\int_{-t_0}^{-\tau_-} dt 
\sum_{i=1}^2 [\tilde \flambda_i(t)-\tilde \blambda_i(t)] 
+
\log {\sin \btheta(t_0) \over \sin \ftheta(t_0)},
\label{trans}
\end{equation}
where $\sin \ftheta(t_0)$ is the four dimensional volume 
spanned by the vectors $\fe_i(t_0)$ and $\rev \tilde \fe_i(t_0)$
$(i=1,2)$, and  $\sin \btheta(t_0)$ is defined similarly.
The quantity (\ref{trans}) can be calculated
by using the forward evolution of the semi-unstable space, because
$\tilde \blambda_i(t)$ and $\tilde \be_i(t)$ are identical to
$\slambda_i(\tilde \Gamma(-t);f_0)$ and 
$\se_i(\tilde \Gamma(-t);f_0)$.


In this way, we can calculate the numerical value of $I$ for
each trajectory and check whether or not $\bra \exp(-I) \ket_{\rm mc}$
is unity. However, there is a slight complication here; 
trajectories  with large negative $I$ are rare, but they 
contribute greatly to $\bra \exp(-I) \ket_{\rm mc}$.
Thus, in order to avoid large numerical error in the computation
of  $\bra \exp(-I) \ket_{\rm mc}$ that may result from the inclusion
of too many such rare trajectories, we divided the range of values of 
$I$ into discrete intervals, and removed from consideration all trajectories
whose values of $I$ fell inside such intervals that contained fewer than
a certain minimum number of data points. 
In this way, we obtained  the normalized distribution function $\Pi(I)$,
which is plotted in Fig. \ref{fig1}. Using the data contained herein, 
for example, we obtain $\bra \exp(-I) \ket_{\rm mc}=1.03$.

However, the value obtained for $\bra \exp(-I) \ket_{\rm mc}$
in this manner depends on how we divide the range of values of $I$ 
into intervals. (In fact, without changing the shape of $\Pi(I)$ 
shown in Fig. \ref{fig1} significantly,  we can make 
$\bra \exp(-I) \ket_{\rm mc}$ identically equal to 1.)
It is thus necessary to find a better method to determine
the value of $\bra \exp(-I) \ket_{\rm mc}$ from our data.
One such method is to consider the asymmetry around the peak 
value of the function $\Pi(I)$. As seen in Fig. \ref{fig2}, it
is expected that the equality 
\begin{equation}
\log \Pi(I)-\log \Pi(-I)=I
\label{main2}
\end{equation}
holds. That is, $\Pi(I)$ should possess the  symmetry described
by the fluctuation theorem \cite{Evans,Gal,Kur,Maes}. 
Then, since (\ref{main}) can be derived from (\ref{main2}), and 
since the situation depicted in fig. {\ref{fig2} seems to be 
independent of the manner in which we treat the data, 
we  conclude that  (\ref{main}) holds.


Finally, we present a theoretical argument for (\ref{main}). 
We note, however, that this argument lacks mathematical rigor. 

Let us consider $M$ trajectories  whose initial 
conditions at $t=\tau_-$ are sampled from the microcanonical ensemble,
where $M$ is a sufficiently large number. Here, in order to
simplify the theoretical arguments, we consider the microcanonical 
ensemble to be defined by the uniform measure on 
the region between two energy surfaces with energies $E_0$ and 
$E_0+\delta E$, where $\delta E \ll E_0$.  

We divide the phase space into small cells $\{ \Delta_i \}_{i=1}^\Omega$,
each of which has  volume  $\epsilon$. We also discretize time 
as $t_n=n(\tau_+-\tau_-)/N+\tau_-$, where $0 \le n \le N$.  
We can  characterize  a trajectory by a set of integers  
$(i_0,\cdots, i_N)$, where the phase space point on 
the trajectory at time $t_n$ is included in  the cell 
$\Delta_{i_n}$.  
We then define  the  path  probability $P(i_0,\cdots, i_N)$
as the ratio of trajectories characterized by $(i_0,\cdots, i_N)$ 
on the set of $M$.

The probability $p(i_n,t_n)$ of finding the phase space point of 
an arbitrary one of the $M$ trajectories in the cell 
$\Delta_{i_n}$ at time $t_n$, is given  the sum of 
$P(i_0,\cdots, i_N)$ over all configurations 
$(i_1,\cdots,i_{n-1},i_{n+1},\cdots,i_N)$. We note that
$p(i_n,t_n)$ can be interpreted as the time evolution of the 
distribution function for $M$ phase space points which
are sampled from the microcanonical ensemble at $t=\tau_-$.

When $\tau_+-\tau_-$ is sufficiently large, 
phase space points on the $M$ trajectories at time $t_N$
may spread over the semi-unstable space in  the $i_N$th cell,
subject to the relation $p(i_N,t_N) \not =1$. Therefore, it is expected that
$P(i_0,\cdots, i_N)$ can be expressed as
\begin{equation}
P(i_0,\cdots, i_N)=\bLambda(i_0,\cdots, i_N)p(i_N,t_N),
\label{nm}
\end{equation}
where $\bLambda(i_0,\cdots, i_N)$ is the probability of finding 
an arbitrary one of the trajectories that are in the semi-unstable 
space in the $i_N$th cell at time $t=t_N$ to be characterized by 
$(i_0,\cdots, i_N)$.
Then we introduce the similarly defined quantity
$\fLambda(i_0,\cdots, i_N)$ as the probability of finding an arbitrary 
one of the trajectories that are in the semi-unstable space
in the $i_0$th cell at time $t=t_0$ to be characterized by $(i_0,\cdots, i_N)$.
Then, using (\ref{nm}), we have the trivial identity
\begin{equation}
\sum_{(i_0,\cdots, i_N)} P(i_0,\cdots, i_N) 
{\fLambda(i_0,\cdots, i_N)p(i_0,t_0) \over
 \bLambda(i_0,\cdots, i_N)p(i_N,t_N)}  
=1.
\label{triv}
\end{equation}
Now, we conjecture that there is an appropriate limit 
in which $\fLambda(i_0,\cdots, i_N)$ and $\bLambda(i_0,\cdots, i_N)$
approach  $\exp(-\int_{\tau_-}^{\tau_+}\sum_{i=1}^2 \flambda_i(t) )$
and  $\exp(-\int_{\tau_-}^{\tau_+}\sum_{i=1}^2 \blambda_i(t) ) $,
respectively. 
Also, from  Liouville's theorem, we know that $p(i_0,t_0)=p(i_N,t_N)$ 
holds in the limit $\epsilon \rightarrow 0$.  Therefore, it is 
reasonable to expect 
that (\ref{main}) can be derived from (\ref{triv}). 


In summary, we have presented numerical evidence for the validity of
the non-equilibrium  equality (\ref{main}) together with a theoretical 
argument.

In a previous paper, by numerically studying  a Fermi-Pasta-Ulam model 
with a time-dependent nonlinear term \cite{SK1}, 
we have found that the Boltzmann entropy
difference has a certain relation to the excess information loss
$H_{\rm ex}$. 
In the billiard model we study presently, $H_{\rm ex}$
is given by the time integration of $\flambda_1(t)-\slambda_1(\Gamma(t);f_1)$
over the region $t > t_0$. Therefore, $I$ is expected to be related 
to the Boltzmann entropy difference in the thermodynamic limit.
Indeed, it has been conjectured that  these quantities are equal 
in the thermodynamic limit\cite {SK2}. Direct numerical evidence 
supporting this conjecture will be presented elsewhere, 
together with a complete theory.

Our study is apparently related to the fluctuation 
theorem, as indicated by the fact that our data apparently satisfy
(\ref{main2}). However, it should be pointed out  that the entropy 
production ratio in thermostatted models is given by  the phase space 
volume contraction ratio, which is related to 
$\flambda_i- \tilde \flambda_i$, not $\flambda_i-\blambda_i$.
Therefore, in extending our analysis to non-equilibrium steady states,
the irreversible information loss defined by (\ref{Idef}) does 
{\it not} become the entropy production. Rather, we believe that this
comes to represent a certain quantity related to  steady state 
thermodynamics, which was proposed phenomenologically by 
Oono and Paniconi\cite{Oono}.


The author acknowledges H. Tasaki for critical discussions on (\ref{main}), 
and T. S. Komatsu for discussions on the model we studied numerically. 
He also thanks G. C. Paquette for critical reading of the manuscript
and for his comments on the theoretical ideas presented here.
This work was supported by grants from the Ministry of Education, Science, 
Sports and Culture of Japan, Nos. 12834005 and  11CE2006.


\begin{figure}
\centerline{\psbox{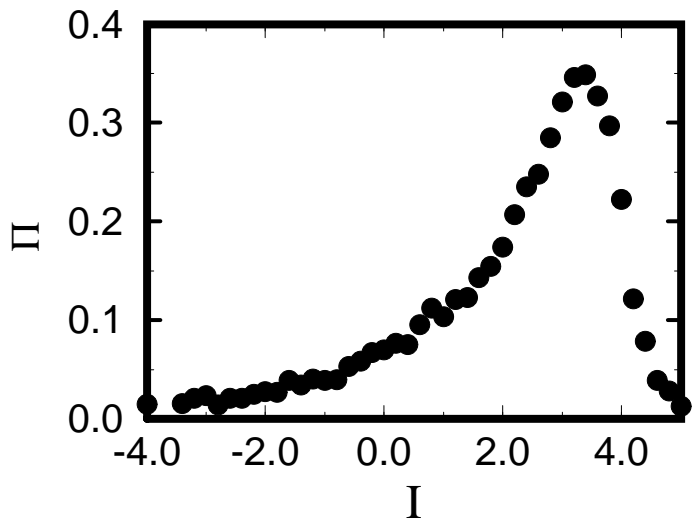}}
\caption{
The normalized distribution function $\Pi(I)$.
The values of $E_0$, $f_0$, $f_1$, $\eta$  and  $\tau_\pm-t_0$
are here $1.0$, $0$, $0.5$, $0.02$  and $\pm 400$,
respectively. We construct a histogram
by dividing the interval $[-5,5]$ into $50$ boxes for the $9964$ samples
considered and removing boxes in which the number of data points 
is less than $25$.
}
\label{fig1}
\end{figure}

\begin{figure}
\centerline{\psbox[scale=1.0]{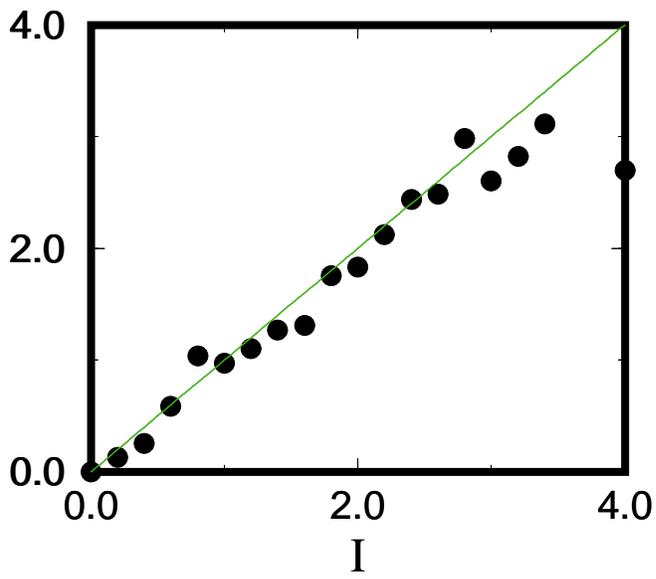}}
\caption{
$\log \Pi(I)-\log \Pi(-I)$. The dotted line corresponds to 
$\log \Pi(I)-\log \Pi(-I)=I$.}
\label{fig2}
\end{figure}

\end{document}